\documentstyle[aps,preprint]{revtex}
\date{\today}
\begin{document}
\title{\bf Degeneracy of Multi-Component
Quantum Hall States Satisfying Periodic Boundary Conditions}
\author{I. A. McDonald~\cite{byline}}
\address{
Department of Physics, Princeton University\\
Princeton, New Jersey 08540}
\maketitle
\begin{abstract}
In systems
subject to periodic boundary conditions,
Haldane has shown that states at arbitrary filling fraction
possess a degeneracy with respect to center of mass translations.
An analysis is carried out for multi-component electron systems
and extra degeneracies are shown to exist.
Their application to numerical
studies is discussed.
\end{abstract}

\def\prl{{\sl Phys.\ Rev.\ Lett.\ }}
\def\pra{{\sl Phys.\ Rev.\ A\ }}
\def\prb{{\sl Phys.\ Rev.\ B\ }}
\def\prc{{\sl Phys.\ Rev.\ C\ }}
\def\prd{{\sl Phys.\ Rev.\ D\ }}
\newpage
\section{Introduction}

Following Laughlin's original work~\cite{laughb},
Halperin~\cite{halpb} proposed an extension to the
original Laughlin-Jastrow wavefunction incorporating the
possibility that the electrons have some further degree
of freedom in addition to their two-dimensional
coordinate, which in Halperin's original observation
was taken to be spin.
Recent experiments~\cite{eisen} on electron systems where the
extra degree of freedom can be seen to indicate
the state of the electron in the third direction
seem to suggest the existence of new universality classes
of states, as first proposed by Haldane and Rezayi~\cite{halde}
and analyzed by Yoshioka, MacDonald, and Girvin~\cite{yosh}.
The possibilities raised by this extra degree of
freedom have been analyzed in various
contexts~\cite{he,hea,wena,wenb,wenc,wend,wilcz,macda,macdb,macdc,iana,ianc}.

In the standard treatment of the problem of electrons
in a magnetic field subject to periodic boundary
conditions (PBCs), Haldane~\cite{haldd} has shown that every
eigenstate of a translationally invariant Hamiltonian
has a degeneracy with respect to
center of mass translations.  For a state at filling fraction
$p/q$ the degeneracy is simply $q$ fold.
Haldane and Rezayi~\cite{haldc} further constructed
the explicit generalization of
Laughlin's wavefunction to the PBCs,
again showing the $q$ fold degeneracy inherent to
states in this geometry.
In light of the interesting possibilities raised
by the double layer systems, it is natural to
generalize the wavefunction construction to
multi-component systems
subject to periodic boundary conditions.

The remainder of the paper is organized as follows.
In the next section we review the construction of
the single layer generalization of Laughlin's wavefunction
subject to periodic boundary conditions.
We then extend this construction to
multi-component systems,
showing interesting new degeneracies
distinctive to these systems.
Finally, we discuss the quantum numbers
of these states with regard to numerical
studies and their use in distinguishing
between possible ground states.

\section{Laughlin Wavefunction in PBC}
We shall construct the Laughlin state subject to
periodic boundary conditions (PBC's), closely
following~\cite{haldc}.
In a system subject to PBCs
in a magnetic field, the
translation operator takes the form
\begin{equation}
t({\bf a})= {\rm exp} \bigl({\bf a} \cdot
({\bf \nabla} - {ie \over \hbar}{\bf A})
- i{{\bf a} \times {\bf r}\over l^2} \bigr)
\label{trans}
\end{equation}
and obeys the non-commutative algebra
\begin{equation}
t({\bf a}) t({\bf a}^{\prime})
= t({\bf a} + {\bf a}^{\prime})
e^{i {{\bf a} \times {\bf a}^{\prime} \over 2 l^2}}
\end{equation}
where
$l = \sqrt{(\hbar /e B)}$ is the magnetic length.
We wish to impose generalized boundary conditions
by requiring that all physical quantities
be invariant under translation of
any particle by the set of
translations
${\bf L}_{mn}= m {\bf L}_1 + n {\bf L}_2$
where
\begin{equation}
\vert {\bf L}_1 \times {\bf L}_2 \vert
= 2 \pi N_{\phi} l^2
\label{fluxnu}
\end{equation}
and
$N_{\phi}$ is the number of flux quanta.
We impose the general boundary conditions on
the wavefunction for any particle $i$
\begin{equation}
t_i({\bf L}_{mn})\Psi = (\eta_{mn})^{N_{\phi}}
e^{i {\bf \Phi}_0 \cdot
{\bf L}_{mn}}\Psi
\label{pbc}
\end{equation}
where
$\eta_{mn} = (-1)^{(m+n+mn)}$
and we will choose
${\bf \Phi}_0 = 0$ as our boundary condition.
The physical region
under consideration can be seen to be defined by four
points $z = {1 \over 2} L_1(\pm 1 \pm \tau )$
where $\tau = {L_2 \over L_1} e^{i \theta}$.
We shall use the
symmetric gauge ${\bf A} = (-By/2,Bx/2).$
The single particle wavefunctions in
the lowest Landau level are then given by
\begin{equation}
\psi(x,y) = e^{-{1 \over 4l^2}
{\vert z \vert}^2} f(z) {\rm \ \ \ \ \ } z=x+iy
\label{wf}
\end{equation}
where $f(z)$ is an analytic function
with $N_{\phi}$ zeros in the principal
region.
Applying the boundary conditions to~(\ref{wf})
we find
\begin{equation}
{f(z+L_{mn}) \over f(z)} = (\eta_{mn})^{N_{\phi}}
{\rm exp} \biggl( {L_{mn}^{*} L_{mn} + 2L_{mn}^{*} z
\over 4 l^2} \biggr )
\label{first}
\end{equation}
where we write a two-dimensional vector
$\bf a$ as $a_x + i a_y.$
The basic building block
that we shall use is the quasi periodic
function $w(z)$
which obeys
\begin{equation}
{w(z+L_{mn}) \over w(z)}
=\eta_{mn} {\rm exp} \bigl( {2 L_{mn}^{*} z
+ L_{mn}^{*} L_{mn} \over 4 N_{\phi} l^2} \bigr ).
\label{spw}
\end{equation}
An explicit representation of the function $w(z)$
is given by
\begin{equation}
w(z) = {\rm exp} \biggl(
{z^2 \over 4 N_{\phi} l^2} \biggr)
\Theta_1( \kappa z \vert \tau)
\label{wex}
\end{equation}
where
$L_{mn} = \kappa^{-1} \pi (m + n \tau)$
and $\Theta_1( \kappa z \vert \tau)$ is the
odd elliptic theta function.
We can therefore construct the single particle
wavefunctions
\begin{equation}
\Psi_{\{ a_{\alpha} \} }({\bf r})
= \biggl( \prod_{\alpha=1}^{N_{\phi}}
\varphi({\bf r} - {\bf a_{\alpha}})
{\rm exp} \bigl ({i \bar {\bf a}
\times {\bf r} \over 2 N_{\phi} l^2} \bigr) \biggr)
\end{equation}
where
\begin{equation}
\bar {\bf a} = {\sum_{\alpha}
{\bf a_{\alpha}} \over N_{\phi}}
\end{equation}
and
\begin{equation}
\varphi({\bf r}) = w(z) {\rm exp} \biggl[
- \biggr({z^* z \over 4 N_{\phi} l^2}
\biggl) \biggr]
\label{build}
\end{equation}

The number of linearly independent solutions
of~(\ref{first}) and can be discerned in the following way.
Inserting the above form into the boundary conditions
yields the constraint
\begin{equation}
\bar {\bf a} = {{\bf L}_{pr} \over N_{\phi}}
\label{coo}
\end{equation}
where
${\bf L}_{pr}$ is restricted to be
a primitive translation.
There are therefore $(N_{\phi})^2$ possible
values for $\bar {\bf a}.$
One way in which to resolve the degeneracy is
to form a superposition of states where every
zero is shifted by the same amount in some
primitive direction.
This is equivalent to constraining the
wavefunction to be invariant under a
translation of the electron coordinate in this
direction.
Given one set of zeros $\{ a_{\alpha} \}$
which satisfies the boundary condition~(\ref{coo}),
another set can be generated by
shifting the zeros
\begin{equation}
a_{\alpha}
\rightarrow  a_{\alpha} + {{\bf L}^{\prime}_{pr}
\over N_{\phi}}
\label{cot}
\end{equation}
where ${\bf L}^{\prime}_{pr}$ is
also a primitive translation.
Let us then form a linear superposition of states with
shifted zeros
\begin{equation}
\Psi({\bf r}) = \sum_{\gamma = 1}^{N_{\phi}}
\Psi({\bf r})_{\{a_{\alpha,\gamma} \}}
\label{e}
\end{equation}
where
\begin{equation}
a_{\alpha,\gamma+1} = a_{\alpha,\gamma} + {K_1
\over N_{\phi}}
\end{equation}
where $K_1$ is a primitive translation.
We can operate on the above
wavefunction by performing an
overall shift of the zeros
\begin{equation}
a_{\alpha,\gamma}^{\prime}
= a_{\alpha,\gamma} + n_k {K_2 \over N_{\phi}}
\end{equation}
where $K_2$ is a
primitive translation obeying
\begin{equation}
\vert {\bf K}_1 \times {\bf K}_2
\vert
= 2 \pi N_{\phi} l^2.
\end{equation}
We can therefore see that
$n_k=1,...,N_{\phi}$ using~(\ref{coo}).
The number of linearly independent solutions
is given by $N_{\phi}$.
It is important to note that this is one program
which generates a space of linearly independent
solutions but not the only one.

We now consider the many particle Laughlin wavefunction
at Landau level filling
$\nu = p/q = 1/m$ on the plane
\begin{equation}
\Psi (\{z_i\}) = \prod_{i<j}
(z_i-z_j)^m \prod_i {\rm exp} \biggl[-\biggl(
{z^*_i z_i \over 4 l^2} \biggr)\biggr].
\label{wavef}
\end{equation}
In the following, we shall denote
$N_e = p \bar N$ and $N_{\phi} = q \bar N.$
When the system is subjected to PBCs,
this wavefunction
generalizes to
\begin{eqnarray}
\Psi(\{ r_i \}) &=& \bigl( \prod_{i<j} \varphi({\bf r_i}
- {\bf r_j})^m \bigl) \nonumber\\
&\times & \prod_{\alpha=1}^{m} \biggl( \varphi({\bf R}
- \bar N {\bf a}_{\alpha})
{\rm exp} \bigl( {i \bar {\bf a} \times
{\bf R} \over 2 N_{\phi} l^2}
\bigr) \biggr)
\label{wavet}
\end{eqnarray}
where
\begin{equation}
{\bf R} = \sum_i {\bf r_i}
\end{equation}
and $\varphi({\bf r})$ is defined in~(\ref{build}).
If we apply the translation operator to any of the coordinates
and apply the boundary conditions,
we find
\begin{equation}
\bar {\bf a} = {\sum_{\alpha} {\bf a}_{\alpha} \over m}
= {{\bf L}_{pr} \over m}
\end{equation}
where ${\bf L}_{pr}$ is restricted to
be a primitive translation.
The center of mass portion of the wavefunction
can be seen to be formally equivalent to
the single particle wavefunction for a
particle of charge $e N_e$ which
sees $m$ flux quanta.
By analogy with the single particle case, we
therefore conclude that the Laughlin state on
the torus has an $m$ fold degeneracy related
to the action of the center of mass translation
operator.
Very generally, this degeneracy is inherent to
every eigenstate, as was shown by Haldane~\cite{haldd}.

\section{Extension to Multi-Component Systems}
We can write the planar extension of Laughlin's
original wavefunction to a system with
$N_{\rm s}$ species of electrons as
first expressed by Halperin~\cite{halpb}
in the two-component case
\begin{equation}
\Psi^{K} [\{z_i\}] ~=~
\prod_{i<j} (z_i - z_j)^{K(\sigma_i,\sigma_j)}
\prod_{i} \exp\bigl(-{1\over 4l^2} \vert z_i \vert^2\bigr)
\label{layera}
\end{equation}
where $\sigma_i$ is the species quantum number
and $K$ is an $N_{\rm s} \times N_{\rm s}$ symmetric
matrix encoding the correlations
between the electrons
where we impose
${\rm Det}K > 0.$
If we consider the plasma analogy, we
find that in order to have a uniform
fluid stable against fluctuations,
we must choose
${\rm Det}K \geq 0.$
We shall consider the case of
${\rm Det}K = 0$ separately.
The matrix $K$ also makes an appearance
in effective theories of the fractional
quantum Hall effect, as the long
distance physics of the Hall fluid can
be described by the Lagrangian
\begin{eqnarray}
{\cal L} = {1 \over 4 \pi} \bigl(
\sum_{\sigma,\sigma^{\prime}}
K(\sigma,\sigma^{\prime}) \epsilon^{\mu \nu \lambda}
a_{\mu}^{\sigma} \partial_{\nu}
a_{\lambda}^{\sigma^{\prime}}
&+& 2 \epsilon^{\mu \nu \lambda}
\sum_{\sigma} t^{\sigma} A_{\mu}
\partial_{\nu} a_{\lambda}^{\sigma}\bigr)
\nonumber \\
&+&{\rm Maxwell \  terms}
\label{lag}
\end{eqnarray}
where $a^{\sigma}$ are $N_s$ Chern-Simons gauge fields.
This Lagrangian has been discussed extensively
previously~\cite{wend}.
The two-component case is of special
physical interest, where the
two species of electron may represent
the two possible physical spin
states of an electron or,
in double layer systems,
which of two layers an electron resides
in.
We have
suppressed the quasi-spin portion of the
wavefunction, which in the case of
fermions would correctly anti-symmetrize the
overall wavefunction.

We write the Halperin multi-component
wavefunction subject to PBCs
as
\begin{eqnarray}
\Psi^{K} [\{z_i\}] &=&
\prod_{i<j} \varphi(z_i - z_j)^{K(\sigma_i,\sigma_j)}
\nonumber\\
& \times & \prod_{\alpha} \biggl(
\varphi \bigl(\sum_i S^{\alpha} (\sigma_i) z_i
-\Gamma_{\alpha} \bigr)\biggr)\nonumber\\
& \times & \prod_{\alpha} \biggl(
{\rm exp} \bigl(i {{\bf \Gamma_{\alpha}}
\times \sum_i S^{\alpha}
(\sigma_i) {\bf r_i}
\over 2 N_{\phi} l^2} \bigr) \biggr)
\label{layerb}
\end{eqnarray}
where $\{ S^{\alpha} (\sigma) \}$
are integers.
We also note the constraint that each electron
must have $N_{\phi}$ zeros in the wavefunction,
implying
\begin{equation}
N_{\phi} = \sum_j K(\sigma_i,\sigma_j)
\end{equation}
which correctly gives us the
filling fraction
\begin{equation}
\nu = \sum_{\sigma,\sigma^{\prime}}
K^{-1}(\sigma,\sigma^{\prime}).
\end{equation}
We can apply the PBCs
to any of the electrons to get the
constraints on the center of mass portion of
the wavefunction.
Applying the translation operators to any
particle and using the boundary
conditions, we find
\begin{equation}
K(\sigma,\sigma^{\prime})
= \sum_{\alpha=1}^{m} S^{\alpha}(\sigma)
S^{\alpha} (\sigma^{\prime})
\ \ \ \ \ m \geq N_{\rm s}.
\label{cons}
\end{equation}
and
\begin{equation}
\sum_{\alpha} \Gamma_{\alpha}
S^{\alpha}(\sigma_i) = L_{mn}(\sigma_i)
\label{constraint}
\end{equation}
where $L_{mn}(\sigma)$ is a primitive translation.
Therefore, given that $S^{\alpha}(\sigma)$ is
integer valued, this implies that the matrix
$K(\sigma,\sigma^{\prime})$ must be
positive definite, as is necessary for
thermodynamic stability.
We can also impose the constraint on the
lowest common denominator
\begin{equation}
{\rm l.c.d} \{ S^{\alpha}(\sigma_1),
S^{\alpha}(\sigma_2),...,S^{\alpha}
(\sigma_{N_s}) \}
= 1
\end{equation}
for all values of $\alpha.$

We now will determine the number
of linearly independent solutions
to these equations, or
equivalently, the
number of independent
ways of arranging the zeros of
the center of mass wavefunction.
For the purposes of forming
sets of zeros such that the corresponding
wavefunctions are orthogonal,
we may constrain the
center of mass portion of the wavefunction
to be invariant under
independent translations of the individual
centers of mass of each species,
keeping in mind that we are not constraining the
overall wavefunction to be invariant under these
translations, only the center of mass wavefunction
independent of the relative piece.
In order to achieve this, we will
form a superposition of states with shifted zeros
as we did in the single particle case.
We then wish to determine the
set of allowed translations $\{ \gamma_j \}$
of the zeros such that
\begin{equation}
\Gamma_{\alpha} \rightarrow \Gamma_{\alpha}
+\sum_j S^{\alpha} (\sigma_j) \gamma_j
\label{altrans}
\end{equation}
is allowed by the boundary conditions.
Inserting this into the boundary conditions,
we find
\begin{equation}
\gamma_j = \sum_k K^{-1}
(\sigma_j,\sigma_k) L_{mn}(\sigma_k).
\label{tran}
\end{equation}
By analogy with the
single particle Hilbert space,
the number of linearly independent solutions
is simply given by
\begin{equation}
{\rm Deg} = {\rm Det} K.
\label{res}
\end{equation}
As an example
we consider
the two component case
where the $K$ matrix is given by
\begin{equation}
K = \left(\matrix{m_1&n\cr
                  n&m_2\cr}\right)
\end{equation}
and the degeneracy is
$m_1 m_2 - n^2.$
This result was noted in the context of
Chern-Simons effective field theory approaches to
the Hall effect~\cite{wenc,wilcz} but not elaborated
upon.  The explicit result is crucial
in numerical simulations, if one is to
try to distinguish between possible
ground states.  The degeneracy provides another
quantum number in addition to filling fraction
to determine the universality class
of a given incompressible quantum fluid.
It is important to make several comments.

1.  Every eigenstate of a Hamiltonian that depends
only on
the relative position of the particles has a degeneracy of
$q$ on the torus if the filling fraction is given by $p/q.$
The degeneracy described here is only a feature of the
state given by the Laughlin-Halperin form~(\ref{layera}).
The overall multi-component degeneracy is due
to the independent
translation of the zeros of the centers-of-mass
of different species,
subject to constraints from the correlated part
of the wavefunction.
This includes $q$ translations of the
overall center of mass of the system,
as well as
translations of the zeros
which can be interpreted
as relative translations of
the centers-of-mass of
the different species.

2. We have assumed in the derivation that
${\rm det} K > 0.$
If ${\rm det} K =0$
and $K(\sigma,\sigma^{\prime}) = m$
the system is
equivalent to a single layer $\nu = 1/m$
Laughlin state with an extra
degree of freedom.
In this case, the state has
the usual center of mass degeneracy of $m.$
In addition, the system has a residual $U(1)$
degeneracy with respect to the number of
particles of each species.  While the
total number of particles is conserved,
the Hamiltonian is invariant under rearrangements
of the number of particles of each species.
This wavefunction has many other interesting features,
as has been investigated in several papers~\cite{wena,macda,macdc}.

\section{Quantum Numbers of Multi-Component States}
In his analysis of two-dimensional electron
systems in magnetic fields subject to PBCs,
Haldane~\cite{haldd} constructed a general symmetry
formalism which clarified the center-of-mass
degeneracy seen previously in finite size studies
subject to PBCs.
This construction has proven to be very useful
in finite size studies on the torus,
simplifying the calculations
as well as providing a correct classification
of states allowing a direct comparison
with studies done in other geometries.
We shall briefly review this analysis
and proceed with its application to
the multi-component systems.

As before
we shall consider a physical
region defined by the
vectors ${\bf L}_1$ and ${\bf L}_2$ and
require physical quantities to
be invariant under translation
through ${\bf L}_{mn}$ where
${\bf L}_{mn} = m {\bf L}_1 + n {\bf L}_2.$
We can consider the operator that generates
relative translations between the particles
while maintaining the center of mass, defined to be
\begin{equation}
\tilde t_i(p {\bf L}_{mn})
= \prod_j t_i({p {\bf L}_{mn} \over N_e})
t_j({-p {\bf L}_{mn} \over N_e})
\label{topo}
\end{equation}
where $N_e$ is the number of electrons in the system
and the filling fraction is given by
$\nu = {N_e / N_{\phi}} = {p / q}$
and the operators $t_i({\bf a})$ are the
translation operators acting on
particle $i$ defined as before~(\ref{trans}).
We can rewrite the above operator, using the
fact that the physical states that this operator
acts upon are subject to the conditions~(\ref{pbc}) as
\begin{eqnarray}
\tilde t_i(p {\bf L}_{mn}) &=& t_i(p {\bf L}_{mn})
\prod_j t_j({-p {\bf L}_{mn} \over N_e})
\nonumber \\
&=& (\eta_{mn})^{p N_{\phi}}
{\rm exp} \bigl( i p({\bf \Phi}_0
\cdot {\bf L}_{mn}) \bigr)
T\bigl(-{{\bf L}_{mn} \over \bar N} \bigr)
\label{topt}
\end{eqnarray}
where
\begin{equation}
T({\bf a}) = \prod_i t_i({\bf a})
\end{equation}
is the center of mass operator.
We can therefore classify the eigenstates of a
translationally invariant Hamiltonian
obeying $[H,T({\bf a})]=0$
by the quantum number $\bf k$ defined to be
\begin{equation}
T\bigl({{\bf L}_{mn} \over \bar N} \bigr) \vert \Psi \rangle
= (\eta_{mn})^{pq}
{\rm exp}
\biggl( i {({\bf k} + N_e {\bf \Phi}_0) \cdot {\bf L}_{mn}
\over \bar N} \biggr )\vert \Psi \rangle.
\label{topth}
\end{equation}
As this operator commutes with the
center of mass translation operator
$T({{\bf L}_{mn} \over N_{\phi}})$,
each eigenstate of a translationally
invariant Hamiltonian, labeled by ${\bf k}$,
will have $q$ fold center of mass degeneracy.
In the following we shall specialize
to the boundary conditions with ${\bf \Phi}_0 = 0.$

We can explicitly define eigenstates for the
above operator, useful for
numerical work, in the following way.
We can denote a fundamental
set of translations as
\begin{equation}
t({L_1 \over N_{\phi}})
= t_1 \ \ \ \ \ t({L_2 \over N_{\phi}})=t_2
\end{equation}
and using the relation
\begin{equation}
t_1 t_2 = t_2 t_1 e^{-i {2 \pi \over N_{\phi}}}
\end{equation}
we can define a basis as
\begin{equation}
t_1 \vert m \rangle = e^{-i {2 \pi m \over N_{\phi}}}
\vert m \rangle
\ \ \ \ \ t_2 \vert m \rangle = \vert m+1 \rangle
\end{equation}
where
$\vert m+N_{\phi} \rangle = \vert m \rangle.$
This is the smallest set of translations consistent
with a given set of boundary conditions.
If we denote the many particle occupation number state
as
\begin{equation}
\vert n_1,n_2,...,n_{N_{\phi}} \rangle
\end{equation}
we can write the eigenstate
of~(\ref{topth}) in the following fashion
\begin{equation}
\vert \Psi \rangle = \sum_k
\vert n_1 + qk, n_2+qk,... \rangle
e^{{2 \pi i k \over \bar N} {\tilde j}_y}.
\label{eig}
\end{equation}
If we act with the operator~(\ref{topth}) on this state, we find
\begin{equation}
T\bigl({{\bf L}_{mn} \over \bar N}\bigr)
\vert \Psi \rangle
= (-1)^{m n p q}
e^{-{2 \pi i \over \bar N}
({\tilde j}_x m + {\tilde j}_y n)}
\vert \Psi \rangle
\label{defj}
\end{equation}
where
\begin{equation}
{\tilde j}_x = \sum_{t=1}^{N_{\phi}} n_t \cdot t.
\end{equation}
Therefore, to set up the Hilbert space, we group
sets of states into their ${\tilde j}_x$ value.  We then construct
the full Hilbert space by forming linear superpositions
of these states, with their occupation numbers shifted,
multiplied by a phase factor.  These are our fundamental
basis states.  We note that ${\tilde j}_x$ and
${\tilde j}_y$ are integers
defined
modulo $\bar N.$
If we then define two
reciprocal lattice vectors
as
\begin{equation}
{\bf G}_{\alpha} \cdot {\bf L}_{\beta}
= 2 \pi \delta_{\alpha \beta}
\end{equation}
one can see that the
Brillouin zone pseudomomentum $\bf k$ is
given by
\begin{equation}
{\bf k} = -(j_x {\bf G}_1 + j_y {\bf G}_2)
\end{equation}
where
\begin{eqnarray}
(j_x,j_y) &=&
({\tilde j}_x,{\tilde j}_y) - \biggl({{\bar N}
\over 2},{{\bar N} \over 2} \biggr)
\hspace{.25in} pq = {\rm odd}
\nonumber\\
&=&({\tilde j}_x,{\tilde j}_y)
\hspace{1.1in} pq  = {\rm even}
\end{eqnarray}

The most symmetrical Bravais
lattice that we can consider is
the hexagonal PBCs,
invariant under a
$\pi /3$ rotation.
We have defined the ${\bf k}=0$ point
in~(\ref{topth}) to be the unique point invariant
under a $\pi /3$
rotation, where
$L_1 \rightarrow L_2$ and $L_2 \rightarrow L_2 - L_1.$
Therefore, the invariant point must be the
solution to the equation
\begin{equation}
T\bigl( {{\bf L}_1 \over \bar N}\bigr)
= T\bigl(-{{\bf L}_1 \over \bar N}
\bigr)
= T\bigl({{\bf L}_2 \over \bar N}\bigr)
= T\bigl({{\bf L}_2 -
{\bf L}_1 \over \bar N}\bigr).
\end{equation}
If we separate the last term and solve, we conclude
\begin{equation}
T\bigl({{\bf L}_{mn} \over \bar N}\bigr)
\vert \Psi \{ {\bf k} = 0 \} \rangle
= (\eta_{mn})^{pq}
\vert \Psi \{ {\bf k} = 0 \} \rangle.
\end{equation}
This defines the ${\bf k} = 0$ point.
The significance of these comments for numerical
work is in the fact that in the thermodynamic limit
the ${\bf k}=0$ point becomes rotationally invariant
and therefore the signal for a QHE fluid is
a ${\bf k}=0$ ground state with an energy gap
to any excited states.

We now turn our attention to the multi-component
system.  In this case, the same analysis carries
through as above with the occupation basis being
expanded to include the pseudospin quantum number.
If one acts with the operator~(\ref{topo})
on the single layer wavefunction
at $\nu =1/m$, one finds
the Laughlin state to be a
${\bf k}=0$ eigenstate.
In the following we shall
denote
${\rm Det} K= q N^{\prime}.$
If one acts with the operator~(\ref{topo})
on the wavefunction $\Psi^{K}[\{ z_i \}]$ one
can determine the quantum numbers
\begin{equation}
j_x = p \bar N \sum_i {N_i \over N_e}
p_i \ \ ({\rm mod} \bar N)
\label{acto}
\end{equation}
and
\begin{equation}
j_y = p \bar N \sum_i { N_i \over N_e}
p^{\prime}_i \ \ ({\rm mod} \bar N)
\label{actt}
\end{equation}
where $p_i$ and $p^{\prime}_i$ are
specified integers
and
$N_i$ is the number of electrons
of species $i.$
One can therefore see that in the multi-component systems,
the Laughlin-Halperin like ground state
will have an overall degeneracy given by
${\rm Det} K$
with a
center of mass degeneracy given by $q$
as well as a
$N^{\prime}$ fold degeneracy
related to the different translations
of the centers of
mass of the different species.
There are four points in ${\bf k}$ space that
have parity invariance,
$\{ j_x,j_y \} = \{0,0\},$
which corresponds to
${\bf k}=0$ and three
states that lie on the zone boundaries
$ \{\bar N/2 ,0 \},
\{0,\bar N/2 \}$ and $\{ \bar N/2,\bar N/2 \}.$
The relative portion of the wavefunction
is an eigenstate of parity so the ${\bf k}$
vector of the states representing a Laughlin-Halperin
like ground state can only lie on one of these
four points.
We can therefore write
\begin{equation}
{\rm Det} K= q (N_0 + 3 N_{B})
\label{interes}
\end{equation}
where $N_0$ is the number of ${\bf k}=0$ states
and $N_B$ is the number of
zone boundary multiplets.
There are only
two possibilities
\begin{equation}
{N_e \over N_i p}
\ =\  {\rm even \ integer}
\ \ \ \ \ \ N_0 = N_B = N^{\prime}/4
\end{equation}
for some species $i,$
otherwise
\begin{equation}
N_0 = N^{\prime}, N_B = 0.
\end{equation}
As an example, we examine the $(3,3,1)$ state,
represented by the matrix
\begin{equation}
K = \left(\matrix{3 &1\cr
                  1&3\cr}\right)
\end{equation}
recently shown~\cite{eisen} to be a physically
realizable state.
In this case, as the filling fraction
of the state is $\nu =1/2,$
we have a two fold center of mass degeneracy.
{}From the above analysis, we also find that
there should be degenerate states possessing
$j_x = 0,\bar N/2$ and $j_y=0,\bar N/2.$
There are then four degenerate states
corresponding to $\{ j_x,j_y\}=\{ 0,0\}$,
a ${\bf k}=0$ state,
$\{ \bar N/2,\bar N/2 \},$
$\{ \bar N/2,0 \},$ and $\{ 0,\bar N/2 \}$
for an overall degeneracy of eight as expected.
This result has been
borne out by numerical studies~\cite{ianc}.
It should be noted that this degeneracy is
a feature of the exact $\Psi^{K} [\{ z_i \}]$ state,
which can be generated numerically by
choosing a truncated Hamiltonian possessing
the correct pseudopotentials
to make the $\Psi^{K}[\{ z_i \}]$ state
the unique ground state.

\section{Conclusion}
In this paper it has been shown that the
degeneracy of a multi-component quantum Hall
state on the torus,
denoted in the standard fashion by
the matrix $K$
is given by
${\rm Det} K.$
The quantum numbers of these multi-component states
in Haldane's symmetry analysis have also been
determined.
These predicted quantum numbers provide
a powerful topological invariant with which one
can distinguish possible ground states.
One example of this is in the
$\nu =5/2$ system.  There have been two
wavefunctions suggested to explain the
observed anomaly~\cite{haldf,gre},
both possessing the correct filling fraction.
On the torus, however, in addition to
their center of mass degeneracies,
the Pfaffian state
possesses a three fold degeneracy~\cite{gre} while
the spin-singlet state has a
five fold degeneracy~\cite{haldg}.
While this issue is not yet resolved, it is clear
that the ground state degeneracy is a
useful characteristic
in distinguishing
between these two states numerically.
Lastly, there is a remarkable transition
in the $\nu =2/3$ double layer system
between two different ground states
that occur
at the same filling fraction but with
different degeneracies on the torus,
as seen in numerical studies~\cite{iand}.
It provides a fascinating example of
the importance of degeneracy considerations
in numerical studies on the torus.

We wish to thank F. D. M. Haldane
for valuable discussions and
for all his support.
We also acknowledge financial
support from NSF grant
DMR-92-24077.

\end{document}